# Superconductor modulation circuits for Qubit control at microwave frequencies

Sasan Razmkhah, Ali Bozbey, and Pascal Febvre, *Senior member, IEEE*

*Abstract*— **Readout and control of qubits are limiting factors in scaling quantum computers. An ideal solution is to integrate energy-efficient cryogenic circuits close to the qubits to perform control and pre-processing tasks. With orders of magnitude lower power consumption and hence lower noise, Single Flux Quantum (SFQ) and Adiabatic Quantum Flux Parametron (AQFP) superconductor logic families can reach ultimate performance at cryogenic temperatures. We have created a superconductor-based on-chip function generator to control qubits. The generated signal is modulated up to tens of GHz based on the external input waveform applied to the superconductor mixer stage. This circuit works at 4.2K. A radiofrequency (RF) matching circuit transmits the signal to the ~mK stage after digital amplification and noise reduction.**

## I. Introduction

The rapid development of information and communication technologies enabled new architectures, unconventional algorithms, efficient computing, and green technologies. These developments demand a thorough investigation into new hardware architectures to overcome von Neumann's computing limitations. Many new computing methods have been presented from different disciplines, including novel materials, devices, and architectures. These methods offer solutions for many issues. However, for implementing new algorithms and hardware to emulate quantum phenomena effectively, no other architecture is better than quantum computers themselves. In 2007, the D-Wave Canada-based company declared it was testing a 16-qubit quantum computer called Orion. Later, several companies such as IBM, Google, Microsoft, and Intel announced their quantum computer development projects [1]. Apart from Microsoft, which is working on Majorana-based topological qubits, also based on superconductors, they have focused on qubits made up of superconducting Josephson junctions. For instance, Google used 12 qubits to simulate a simple chemical reaction on a quantum computer [2]. Quantum computers allow studying the physics of electron interactions in different materials which is impossible or unpractical with classical computers [3].

Superconducting qubits are among the most suitable devices to build quantum computers [4]. The creation and development of quantum computer control and readout engineering are crucial to protect qubits from decoherence and to read their physical states. Josephson junction-based quantum bits are at the forefront of qubit developments [5-9].

Scaling quantum computers is also challenging. With about 100 physical qubits, working on different approaches to create logical qubits is possible [10]. Unlike physical qubits in the general case, logical qubits can store quantum data without error. The number of available logical qubits measures the technology maturity towards scaling up and commercialization. With 1000 logical qubits, storing quantum data for about a year should be possible [10]. Scaling up requires downsizing many components, including cables, amplifiers, filters, and electronics. Recently, Google used 53 qubits to examine, depending on the design of the circuit, how a quantum entangled wave can propagate ballistically, and to control the propagation rate of entanglement [10-11].

One of the standing issues in operating qubits is the need for control circuits that do not bring noise or thermal load at ultralow temperatures where quantum phenomena are observed. Therefore, measuring and controlling qubits with room temperature devices is expensive, complicated and cannot easily be scaled to a level of complexity for which quantum computers bring an overwhelming and disruptive advantage. Indeed, to achieve fault-tolerant computation, scaling quantum computers is inevitable. One way to accomplish this is to use cryogenic circuits to reduce the wiring and thermal noise of the control circuit, and to increase the flexibility and reconfigurability of the qubit control and readout interfaces.

This work aims at developing an on-chip superconductor SFQ-based architecture for the readout and control of quantum bits without the need for complex room-temperature electronics. The use of superconductor circuits for qubit control was first theorized by Semenov et al. [13]. Castellano et al. later proposed an on-chip fabrication method [14]. In recent years the growth of the number of teams working in the quantum computing field led to many efforts in this direction. Several works use cryogenic CMOS [14-16], RSFQ [17-21], and AQFP [23] circuits for qubits control and readout.

In this work, we propose an alternative method using SFQ-based logic to generate the needed microwave signal waveform to control the quantum qubits and match the noise level of

Manuscript received; This work was partially supported by the Office of the Director of National Intelligence (ODNI), Intelligence Advanced Research Projects Activity (IARPA), via the U.S. Army Research Office under Grant W911NF-17-1-0120.

(Corresponding author: Sasan Razmkhah.)

Sasan Razmkhah was with Electric an Electronic Engineering, TOBB Economics and Technology University, and currently is with University of Southern California, Ming Hsieh department of Electrical and Computer Engineering, Los Angeles, USA. (e-mail: razmkhah@usc.edu).

Ali Bozbey is with Electric an Electronic Engineering, TOBB Economics and Technology University, Ankara 06560, Turkey (e-mail: bozbey@gmail.com).

Pascal Febvre is with IMEP-LAHC, University Savoie Monte-Blanc, 73376 Le Bourget du Lac, France (e-mail: Pascal.Febvre@univ-smb.fr).

quantum electronics. The proposed circuit has two main advantages over other methods. First it uses SFQ-based digital amplifiers and attenuates the signal before applying it to qubits. This reduces the noise level caused by control electronics since the digital amplifier adds little noise. Second, it uses a JJ as local oscillator followed by a Josephson interferometer which acts as a mixer and modulator, that we call Mix&Mod module. With a JJ-based oscillator, one can apply higher frequencies to drive qubits. The interferometer Mix&Mod module can be controlled with a bias distribution network to modulate the signal appropriately, giving us more control over the circuit.

In the following sections, we discuss the design of Mix&Mod module to generate the required microwave signals for qubit control and the way these devices are connected to the mK stage without introducing thermal noise. Finally, we show simulation results and the circuits fabricated by a commercial foundry.

## II. METHODOLOGY

The superconductivity phenomenon in Josephson qubits and the quantization of magnetic flux in a superconducting ring occur at the macroscopic level, making them good candidates for quantum computing systems. The two close energy levels required for the qubit-based quantum parallel $L$-$C$ oscillator must be separated from other energy levels with sufficient oscillator anharmonicity. This is in practice made by replacing the inductor with a Josephson junction well-known to possess a non-linear Josephson inductance [24] to form Josephson qubits [7-8], [23]. The macroscopic quantum effects predicted by Caldeira-Leggett [25] for small-size Josephson junctions are observed at low temperatures and demonstrated by experimental studies [6]. The energy spectrum of superconducting qubits is obtained by solving the Schrödinger equation for the Josephson-based anharmonic $L$-$C$ oscillator [6]–[8], [26]. Fig 1 shows a simple structure to make qubits using a Josephson junction and a capacitor in a superconductor loop [27].

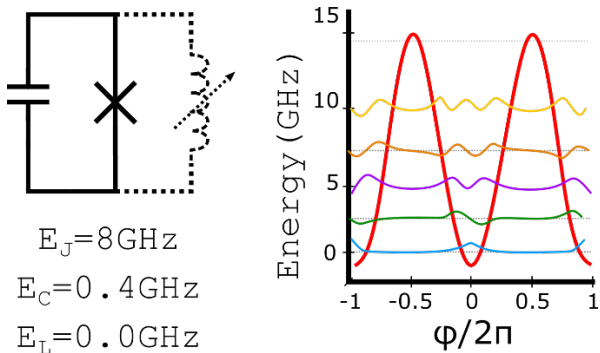

Fig 1 A superconductor loop with a capacitance and a Josephson junction. The non-linear inductance of the JJ and the capacitor form an LC oscillator that acts as a quantum bit. The energy levels show two quantum states for this structure known as the transmon. An inductor can be placed in parallel to the JJ to tune the transmon frequency. This inductor is shown with a dashed line in the schematic.

We must apply the proper microwave signal with specific frequency and duration to control a qubit like a transmon. The signal should be at the resonator's resonance frequency coupled with the qubit, and the waveform of the signal determines which operation (gate) is performed on the qubit.

We use Josephson oscillations as the source for high-frequency signals to reach this goal. Josephson junctions oscillate in the presence of an applied voltage at a frequency of 483.6 GHz/mV, as described in equation 1. In this equation, $\varphi$ is the JJ's phase, $\bar{V}$ is the average or DC voltage over the JJ, $2e$ is Cooper-pair charge, and $\hbar$ is reduced Plank's constant. Fig 2 shows the JJ oscillations as a DC voltage is applied to it. The oscillation frequency increases proportionally with the voltage.

$$\frac{\partial \varphi}{\partial t} = \frac{2e}{\hbar}\bar{V} \stackrel{then}{\Longrightarrow} \varphi = \frac{2e}{\hbar}\bar{V} t + \varphi_0 \qquad (1)$$

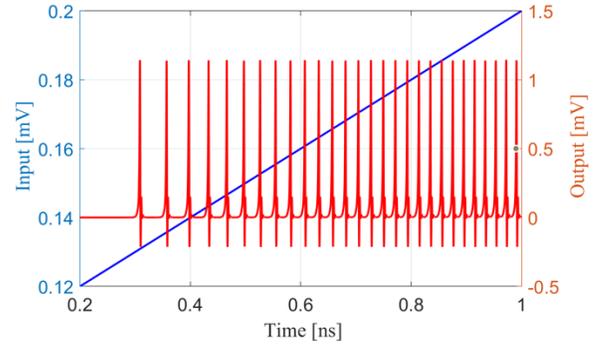

Fig 2 Josephson junction oscillation for a linearly varying DC voltage. As the input DC voltage increases, the JJ oscillates at the Josephson frequency according to equation 1. The output is the time-dependent voltage over the JJ terminals.

The oscillating junction is connected to a Josephson transmission line to reshape the oscillating signal into SFQ pulses suitable for SFQ digital processing. A second stage in the circuit performs the mixing and modulation, of the incoming SFQ signal. The signal at the intermediate frequency (IF) determines the envelope and hence the duration of microwave oscillations; it is brought from outside at room temperature. The mixer output signal is the modulated signal. This signal is amplified by a digital pulse multiplier [28] and then passes through a down-converter and low-pass filter. Fig 3 shows the circuit block diagram for the proposed structure.

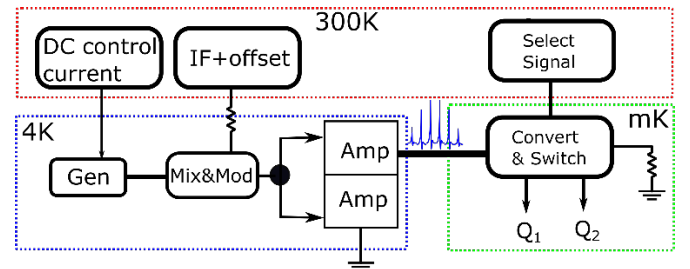

Fig 3 Block diagram of the SFQ control system. Gen is the pulse generator that consists of a JJ and a JTL. Its frequency is controlled with the applied DC bias. Mix&Mod is the signal mixing and modulation stage, and Amps amplify the signal. The interconnect ports and attenuators between the control and the qubit chip are not shown. At mK stage, the Convert & Switch cell matches the frequency to the qubits Q1 and Q2, and switches the control signal between them.

The resulting output signal is then can be distributed by a series of Josephson junctions, controlled by external DC bias lines. The output of the amplifier is connected to the matching ports that match the impedance between the SFQ chip and the mK stage.

Since the SFQ and quantum chips are at 4K (or down to 1K) and mK temperatures, we do not want to carry the thermal noise to the quantum chip stage. Therefore, the modulated SFQ signal is digitally amplified before being sent to the quantum chip and then attenuated afterward to match the quantum noise level [29]. The amplifier does not increase the noise since it multiplies the number of SFQ pulses rather than adding gain in a frequency-domain window.

Another advantage of this structure is its ability to operate at higher frequencies than conventional room-temperature control electronics since it is based on SFQ electronics. Therefore, with the same electronics, one can drive qubits designed for higher frequencies and increase the microwave photon energy to lower the qubits sensitivity to thermal noise.

### III. CONTROL ELECTRONICS DESIGN

The circuits for each part of the system described in section II were designed and simulated using a SPICE-based simulator [30]. Fig 4 shows the circuit schematic of the controller electronics.

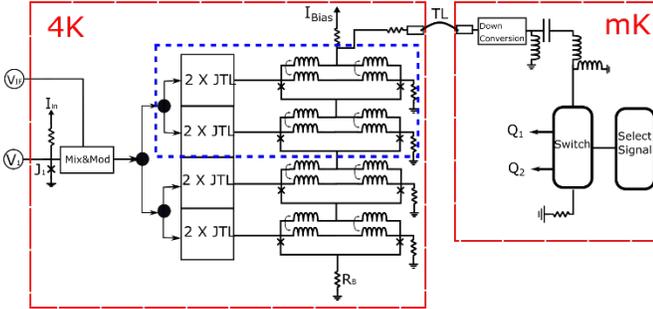

Fig 4 Schematic of the qubit control circuit using SFQ logic. The digital amplifier increases the SFQ pulse amplitude by producing a quantum of flux to each SQUID. The dashed box corresponds to a single amplifier stage. The SQUID output pulses are sent via a matching port through a transmission line (TL) to the mK stage where qubits stand. The Down Conversion cell then matches the frequency of the input signal to the ones of qubits Q1 and Q2. The pulses are filtered to keep only the desired signal, at about 5 GHz. The Switch can change the path of the signal between Q1 and Q2.

The input signal is produced by a biased Josephson junction, whose output signal, shown in Fig 2, is applied to the Mix&Mod module. This module is a multi-junction interferometer consisting of multiple SQUID loops whose critical current is modulated by the magnetic flux present in the loops. A unit cell of the interferometer is a 2-junction-SQUID whose magnetic interference pattern is well known. Fig. 5(a) shows the $k^{th}$ loop of such an interferometer. $L_K$ is the inductance of the loop, $I_K$ and $I_{K+1}$ are the critical currents of the JJs in the loop and $\varphi_K$ and $\varphi_{K+1}$ are their respective phases. The external magnetic field effect on the loop is modeled with the phase source $\varphi_e$, the associated flux is $\phi_{ext}$. The critical current modulation versus external field $\phi_{ext}$ of an interferometer made of four identical JJs, consequently comprising three loops, is shown in Fig 5(b). A N-junction interferometer would be the equivalent of a grating in optics.

The modulation of the JJ critical currents results in the change of the interferometer impedance. Therefore, this circuit can modulate the input signal through the applied field to its loops. This field can be provided by an external IF source, as shown in Fig 3. An offset is used to optimize the response of the interferometer.

The RF signal made of an SFQ pulse train is modulated by the critical current value of the interferometer, which modifies the impedance of the circuit and, consequently, its transmission properties. Increasing the number of junctions of the interferometer makes it even more sensitive, resulting in a modulation that is sharper but also more sensitive to magnetic noise.

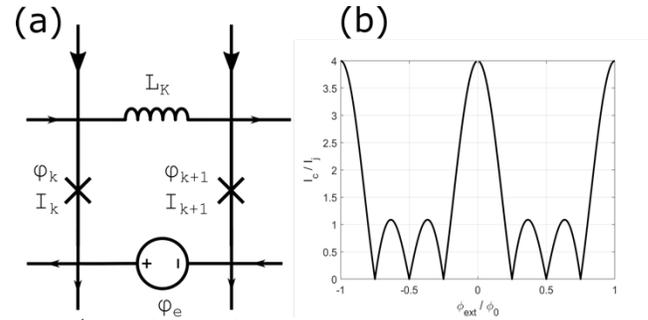

Fig 5 a) $k^{th}$ loop of the interferometric Mix&Mod module that mixes RF and IF signals. b) Normalized critical current versus flux applied to the interferometer, shown for a 4-junction interferometer.

After modulation, the signal goes through the amplifier to increase the energy of the pulses without adding thermal noise to the signal. The amplifier stage is discussed thoroughly in [28]. The digital amplifier stage is necessary for noise matching between different cryogenic levels. It is made of a series of DC SQUIDs inductively coupled to the parallel inductive branches of the circuit. These branches carry pulses cloned by splitters placed at the input of the amplifier (see Fig 4). Each amplifier stage contains two SQUIDs and adds one pulse. Since the pulses are quantized and added together, the noise level does not increase. Therefore, by amplification followed by attenuation, one can average pulses and reduce the noise.

As seen in Fig 6, the digital amplifier performs best with four amplifier stages that consist of eight SQUIDs. In this case, the analog attenuation of the digitally amplified signal can be up to 15dB, which is more than enough to bring the noise down to the quantum level.

The error for the amplifier was measured by giving $10^4$ input pulses and measuring the output pulses at low frequency. The output pulse amplitude should be exactly four times the input pulse value. Otherwise, we consider that there is a digital error. The error vs normalized bias current is shown in Fig. 6. The bias current is only for the SQUIDs as the digital circuits have their own biases. Therefore, while the margin may seem narrow, since this bias is externally applied to the SQUIDs, separate from the rest of the circuit, it does not affect the digital circuit's margin.

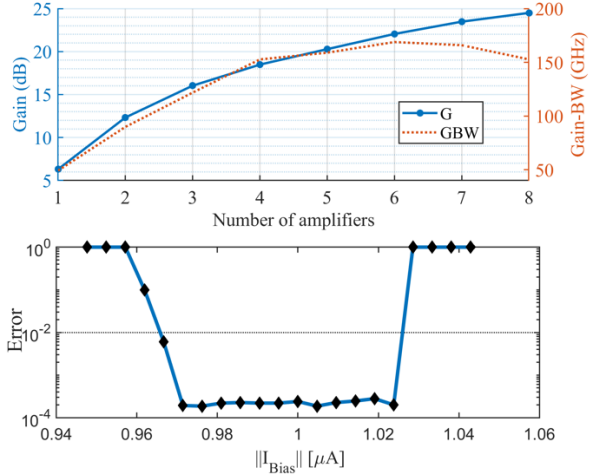

Fig 6 Top: Measured digital amplifier gain (G) and gain-bandwidth (GBW) product. The four-stage amplifier is sufficient to match thermal noise levels after attenuation. Bottom: Error measurement results and bias margin of the amplifier. The bias current value is normalized to the design nominal value.

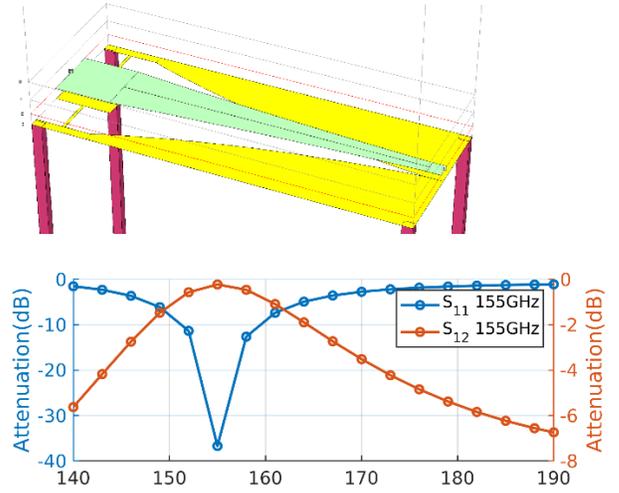

Fig 7 Top: Matching port from microstrip to coplanar waveguide configurations optimized to transmit the SFQ pulse train whose energy is centered at 155GHz. Bottom: Scattering parameters of the matching port.

In a last step a passive bandpass filter is implemented at the amplifier's output to eliminate the signal's DC value and high-frequency noise. The resulting output is a sinusoidal signal whose envelope is determined by the external IF signal. This signal goes to an impedance-matched passive transmission line whose design is based on Mattis-Bardeen calculations [30-31].

The output matching port converts the transmission line low impedance microstrip configuration to a 50Ω coplanar waveguide characteristic impedance. This allows us to pass the signal without significant reflections by coaxial wiring to the mK stage. Fig 7 shows the matching port from the microstrip to the coplanar waveguide for SFQ pulse propagation that needs to be centered at 155GHz. Since a pulse train is directly sent, the frequency of the transmission circuit was set where most of the pulse energy is present [31]. Working at higher frequencies brings the benefit of smaller output port for impedance matching, while the signal is less affected by noise.

Since the control signal needed by the qubits is close from the 5 GHz range, the high frequency pulse train is down-converted on the mK stage to 5 GHz. This is done by a bank of toggle flip-flops (TFF), an SFQ/DC converter and a band-pass filter (BPF). Each TFF cell toggles for every two input pulses, dividing the number of output pulses by a factor of two, resulting with a final signal that is down-sampled. The BPF transforms the down-sampled signal into a sine wave and removes unwanted harmonics and thermal noise.

## IV. SIMULATION RESULTS

After designing each stage, sub-circuits were assembled, and the whole structure was optimized for better parameter margins. For optimization, we used the particle swarm optimization (PSO) method discussed in detail in [33]. Simulations were done with JSIM [30], [34]. The circuits layout were drawn using KLayout software [35]. Inductex [36] was used to extract the parameters from the configuration and match them with circuit schematics values. Simulations were done again with the extracted parameters to ensure the correctness of the design. Microwave simulations for transmission lines were done with SONNET software [37]. For simulating lossy superconductor transmission lines in SONNET, impedances for the fabrication process were extracted using the Mattis-Bardeen formalism [38]. Simulation results for the controller circuit are shown in Fig 8.

Measuring SFQ pulses directly from the amplifier outputs is not practical since the pulses last for a few picoseconds and have an amplitude of about 2 mV, even after amplification. To do measurements at the output stage of amplifiers, we have used SFQ/DC converter circuits which toggle when an SFQ pulse arrives. Therefore, the generated output signal is a non-return-to-zero (NRZ) signal at half the frequency of the pulse train. Thus, to create and measure a 5GHz output signal, the pulse train generated by the Josephson oscillator must be at 10 GHz on the input port of amplifier.

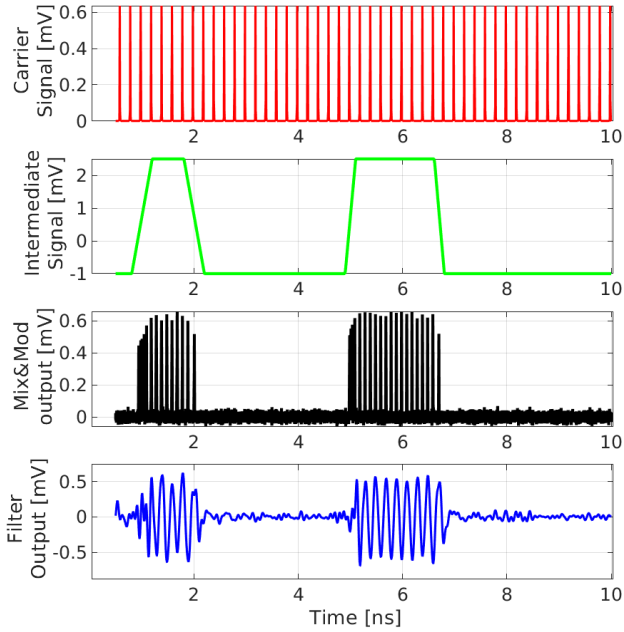

Fig 8 Simulation results for the controller circuit. The carrier signal (top) is generated by the overbiased JJ and then passes through a JTL to fix its shape. The Intermediate Frequency signal (second from top) is externally applied to the Mix&Mod cell, whose output in presence of thermal noise is shown (second from bottom). The output (bottom) is shown after passing through down-conversion, bandpass filtering and attenuation.

## V. FABRICATED CIRCUITS

After designing the layouts, our circuits were fabricated with the HSTP process of the CRAVITY Foundry (currently known as QuFab) [39]. Fig 9 displays the circuits where each part is fabricated separately for testing purposes. Measurement results of the amplifier stage are shown in Fig 6.

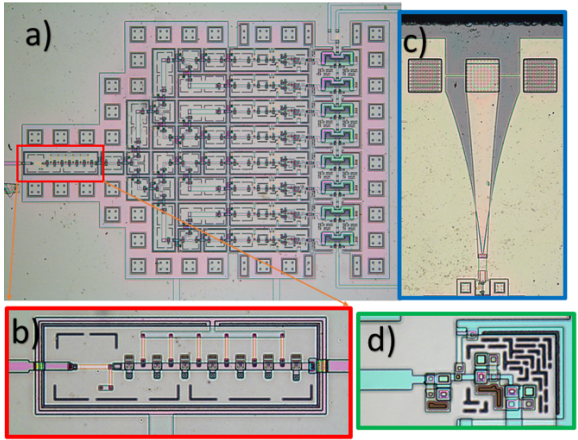

Fig 9 Layout of the fabricated circuit. a) Mixer + amplifier circuit. The optimum number of stages for the amplifier is 4. b) Mix&Mod interferometer circuit. c) Transmission line for off-chip connectors. d) Josephson junction oscillator connected to a JTL.

As mentioned before, we have used SFQ/DC converter cells in the fabricated amplifiers to make direct measurements possible. The Josephson junction oscillator was also tested independently. Results were confirmed by measuring the DC $I$-$V$ characteristic of the junction. The DC voltage is directly correlated to the SFQ pulse-train frequency [40]. See also equation (1) and Fig. 2. The proposed test setup used to ensure the proper operation is shown in Fig 10. The J1 junction generates the RF signal. The IF signal is generated with a function generator and is mixed and amplified on the first chip. The output signal is transmitted to chip 2 (see Fig 10) via matching ports. The received signal is applied to the $J_2$ junction. Depending on the power and frequency of the signal, the $I$-$V$ characteristic of $J_2$ is modified. The whole circuit will be fabricated and tested in the future. The test setup and cryocooler are discussed further in [41].

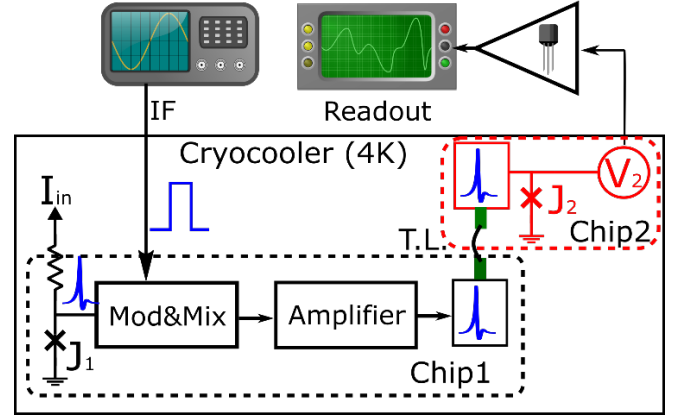

Fig 10 The proposed test setup used to confirm the correct operation of the circuit. TL is the transmission line. In this setup, we emulate sending data from SFQ chip to another chip at cryogenic stage. Chip 1 represents the control circuit and chip 2 is where we want to receive the control signal. By measuring the $I$-$V$ characteristics of the $J_2$ junction one can deduce the frequency and amplitude of the received signal.

## VI. CONCLUSION

We have designed a qubit controller circuit based on SFQ superconductor logic circuits. The circuit consumes less than 50μW and can be used for control quantum gates. Since the controller works nominally at 4K (and down to 1K in fact), the number of cables and RF lines needed for qubits is dramatically decreased. The power consumption is two orders of magnitude lower than for its CMOS counterparts, while the number of JJs is smaller than for other proposed SFQ circuits. The sub-circuits were fabricated separately, the amplifiers were experimentally tested and validated. We expect better noise performance and lower cooling power requirements to enable future scaling of superconductor quantum computers. Another advantage of this structure is the ability to increase the controller's frequency, which will allow more noise-immune qubit operation at higher frequencies with smaller geometries to scale the number of qubits as well, for a given chip area.

ACKNOWLEDGMENT

The circuits were fabricated in the Clean Room for Analog & digital superconductivity, CRAVITY (currently known as QuFab), at the National Institute of Advanced Industrial Science and Technology in Japan. S.R. would like to thank Massoud Pedram for his support.